\documentclass[conference]{IEEEtran}
\IEEEoverridecommandlockouts
\usepackage{cite}

\usepackage{amsmath,amssymb,amsfonts}
\usepackage{algorithmic}
\usepackage{graphicx}
\usepackage{textcomp}
\usepackage{xcolor}
\usepackage[caption=false]{subfig}
\usepackage{multirow}
\usepackage{geometry}
\def\BibTeX{{\rm B\kern-.05em{\sc i\kern-.025emb}\kern-.08em
	T\kern-.1667em\lower.7ex\hbox{E}\kern-.125emX}}

\def\sl{{\text{SL}}}

\def\sync{{\text{sync}}}

\allowdisplaybreaks

\begin{document}

\title{Synchronized SCUBA: D2D Communication for Out-of-Sync Devices
	\thanks{This work was supported by Sierra Wireless Inc. and the Natural Sciences and Engineering Research Council of Canada (NSERC).}
}


\author{\IEEEauthorblockN{Vishnu Rajendran Chandrika$^1$, Gautham Prasad$^{1,2}$, Lutz Lampe$^1$, and Gus Vos$^2$}
	\IEEEauthorblockA{$^1$The University of British Columbia, Vancouver, Canada. \quad    
		$^2$Sierra Wireless Inc., Richmond, Canada. \\
		Email: \{vishnurc@ece, gautham.prasad@alumni, lampe@ece\}.ubc.ca, gvos@sierrawireless.com 
		} 
}

\maketitle

\begin{abstract}
Device-to-device (D2D) communication is an essential component enabling connectivity for the Internet-of-Things (IoT). SCUBA, which stands for Sidelink Communication on Unlicensed Bands, is a novel medium access control protocol that facilitates D2D communications on the sidelink for IoT and machine-type communication (MTC) cellular devices. SCUBA includes support for direct peer-to-peer communication on the unlicensed bands by operating in a time division multiplexed manner to coexist with the underlying primary radio access technology, e.g., long term evolution - MTC (LTE-M). A fundamental requirement in the current version of SCUBA is that the communicating devices are to be synchronized with each other so that timing occasions of the devices can be accurately estimated by each other. However, when the devices are out of sync with each other, which may be caused due to one or more of the devices being out of cellular coverage region or are being served by different base stations, typically observed in mobile devices, operation of legacy SCUBA fails. To this end, we design synchronization methods to establish successful SCUBA links between devices that are out-of-sync with each other. Due to the inherent timing discovery embedded in our method, our solution also extends the operating range of SCUBA by eliminating its reliance on timing-agnostic communication. We  analyze and compare the performance of our methods in terms of power consumption and the resultant impact on device battery life to show the potential of our solutions. 
\end{abstract}

\begin{IEEEkeywords}
IoT, MTC, D2D, sidelink, SCUBA
\end{IEEEkeywords}

\section{Introduction}\label{sec:introduction}
In contrast to conventional cellular communication where user equipments (UEs) communicate with a central base station (BS) on the uplink (UL) and downlink (DL) channels, UEs directly talk to each other on the \textit{sidelink} (SL) in device-to-device (D2D) communications\footnote{We use the terms SL and D2D interchangeably throughout the paper.}. D2D communication is a key component of Internet-of-Things (IoT) and machine-type communication (MTC) networks, as it potentially reduces latency, improves the device battery life, and assists the network in supporting a larger number of connected UEs~\cite{D2Dadv1,D2Dadv2,D2Dadv3,D2Dadv4}. The benefits of using D2D communication can be increased by further transitioning SL transmissions on to the unlicensed frequency bands. This alleviates traffic from the congested licensed cellular bands, while also reducing the associated licensing costs~\cite{andreev2015understanding, liang2018cluster, bouzouita2019estimating}.

As part of reducing the cost and power consumption of IoT and MTC devices, the use of half-duplex frequency division duplexing (HD-FDD) operation is often considered~\cite{hoglund2018overview, BORKAR2020145, sequans_whitepaper,liberg2018mtcbattery, hdfdd_ref}. Applying SL on unlicensed bands for HD-FDD devices using existing commercial D2D protocols such as Bluetooth~\cite{bluetooth_spec}, Zigbee~\cite{zigbee_spec}, and Wi-Fi-Direct~\cite{wifidirect} is expensive, since these radio access technologies require an additional radio chain. Furthermore, these technologies also require manual device pairing and repeated user interventions~\cite{bluetoothuser1,bluetoothuser2,bluetoothuser3}, which reduce the appeal of D2D communications for IoT and MTC applications. Alternative D2D solutions that have been proposed in the literature may counter a few of the above challenges, but still fall short of meeting the needs of incorporating SL communications in low-cost low-power IoT devices. For example, the solutions in~\cite{dtvD2D, wifiD2D1, wifiD2D2, D2DU5G} achieve D2D communication in unlicensed bands, but require significant and continued assistance from a centralized BS for successful operation. This undermines the benefits achievable with the use of SL. Therefore, a novel protocol called SL Communications on Unlicensed Bands (SCUBA) was recently developed for D2D communications in unlicensed bands~\cite{rajendran_journal2020, rajendran2020}. SCUBA operates in a time division multiplexed (TDM) manner with the underlying cellular radio access technology (RAT), e.g., long term evolution - MTC (LTE-M). Due to the TDM nature, SCUBA functions as a secondary RAT and utilizes the existing radio hardware to maintain a single radio architecture in low-cost IoT and MTC devices. Furthermore, it supports operation in the HD-FDD devices and is also upward compatible with other modes of duplexing. Additionally, SCUBA does not require any guidance (e.g., resource allocation) from a centralized BS as is the case with several of the existing D2D communication schemes~\cite{dtvD2D, wifiD2D1, wifiD2D2, D2DU5G}.     


Although SCUBA solves several existing challenges for achieving SL communication on unlicensed bands for low-cost UEs, a fundamental requirement for SCUBA communication is to have the communicating UEs synchronized with each other. This is because a source UE (SRC) communicates with a destination UE (DST) by transmitting SCUBA data on a dedicated time-slot, called the SL paging occasion (SL-PO) of the DST, which it computes using a pre-defined relation~\cite{rajendran_journal2020}. To determine the exact location of the SL-PO in time and to further communicate with each other, the two UEs must be in sync with each other. When both the UEs are in  homogeneous coverage (HC), i.e., being served by the same BS, the devices are perfectly synchronized. However, there are at least three other types of scenarios where this is not the case: 
\begin{enumerate}
    \item Out-of-coverage (OOC): where both devices are out of cellular coverage area
    \item Partial coverage (PC): where one of the UEs is in coverage while the other is not
    \item Coverage-out-of-sync (COOS): where the two UEs are served by two different BSs that are not synchronized with each other.
\end{enumerate}
In terrestrial networks with static BSs, stationary UEs always remain in one of the four coverage scenarios and can therefore use tailored solutions. For example, when two devices are in HC, legacy SCUBA can be successfully used, whereas two UEs in OOC can use other commercial D2D techniques. However, mobile UEs are likely to encounter a different condition at different instances of time. This requires a unified solution that can adapt and operate in any given coverage scenario that could change at any time. To this end, we propose an enhancement to legacy SCUBA, which has already been shown to be the superior D2D RAT among prior arts with respect to power consumption, network latency, and heterogeneous interoperability. Our synchronized SCUBA protocol is compatible with both stationary UEs as well as nodes in an Internet-of-Mobile-Things (IoMT) environment to ensure seamless operation under all of HC, OOC, PC, and COOS conditions. 

Furthermore, we note that due to the lack of synchronization mechanisms and methods to estimate timing advance, the coverage range of legacy SCUBA is restricted by the length of cyclic prefix used. Using LTE-M specifications with an orthogonal frequency division multiplexed (OFDM) sub-carrier spacing of $15$~kHz~\cite{ts_36211}, the maximum range of SCUBA is limited to 1400 meters, which is smaller than typical non-urban macro cell sizes~\cite{tr_36814}. To counter this issue, we present unified  methods that do not only provide synchronization to SCUBA UEs but also extend the range to any arbitrary value that is not limited by time-of-flight.  
In the following, we present our solution along with a review of prior arts and its potential adaptations. We first begin by describing our system model.


\section{System Model}\label{sec:system_model}
\begin{figure}[t]
	\centering
	{\includegraphics[width=7.5cm]{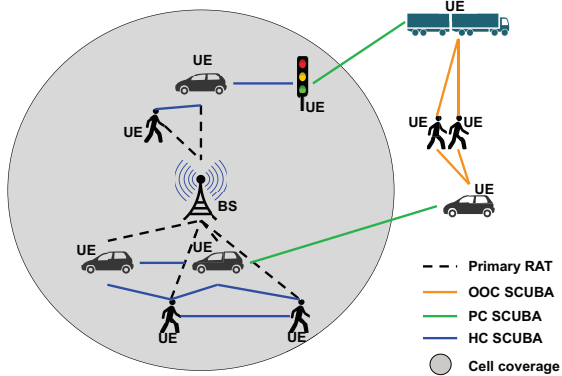}}
	\caption{An illustration of SCUBA coexisting with P-RAT network.}
	\label{fig:ueoperation} \vspace{-.2in}
\end{figure}
\begin{figure}[t]
	\centering
	{\includegraphics[width=7.5cm]{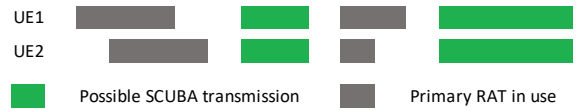}}
	\caption{An illustration of the TDM nature of SCUBA.}
	\label{fig:tdmscuba} \vspace{-.2in}
\end{figure}
We consider a hybrid system model, where the UEs operate in two TDM modes of RATs. The UEs operate an \textit{LTE-like} protocol such as, LTE-MTC (LTE-M), or narrowband-IoT (NB-IoT) as the primary RAT (P-RAT), and SCUBA as the secondary RAT (S-RAT). An illustration of heterogeneous SCUBA network coexisting with P-RAT is shown in Fig.~\ref{fig:ueoperation} for different coverage scenarios such as HC, PC, and OOC, as discussed in Section~\ref{sec:introduction}. While the UEs which are in HC communicate to each other via SCUBA by utilizing the common sync achieved through the cellular coverage, the UEs in OOC, PC, and COOS require new sync methods to have successful communication. Within the devices themselves, a SCUBA SRC transmits to a DST only when both the UEs are free from operation in their P-RATs. Fig.~\ref{fig:tdmscuba} demonstrates this operation to illustrate the TDM nature of SCUBA that allows it to coexist with the P-RAT using the same radio architecture.

\subsection{UE Operation in P-RAT}\label{sec:primary_RAT}
In this subsection, we provide a brief overview of UE operation in the P-RAT with focus on details relevant for SCUBA operation. As an example for clearer illustration, we consider LTE-M as the P-RAT operation on a single radio HD-FDD MTC UE. The UE communicates with the BS via DL and UL sub-frames (SFs) of $1$~ms length with guard periods in between to switch between the transmitter and receiver chains. Due to the sporadic nature of MTC traffic, the UE largely sleeps in its P-RAT and periodically wakes up to sense for any possible incoming messages from the BS. This period is defined by discontinuous reception (DRX) cycles. During the DRX sleep times, the UE is free from P-RAT and is hence available for SCUBA in the S-RAT.

\subsection{UE Operation in S-RAT}\label{sec:secondary_RAT}
S-RAT operation consists of UEs communicating directly with each other using SCUBA messages. Note that any S-RAT communication is possible only when both communicating UEs are free from their respective P-RATs. Consider a pair of UEs, an SRC and a DST. The SRC and DST terminals are not fixed between the two UEs, such that half-duplex bidirectional communication is feasible. Similar to cellular DRX, SCUBA uses SL-DRX cycles to periodically wake the UE up on dedicated time slots called SL-POs to listen for an incoming message, while remaining in the sleep state for the remaining duration of time. The SRC therefore pages the DST on the SL-PO of the latter, which it computes using the unique international mobile subscriber identity (IMSI) and the SL-DRX cycle period of the DST. The IMSI-dependent SL-POs help in reducing potential SCUBA packet collisions. The SRC obtains the IMSI and SL-DRX cycle values of the DST from a central SCUBA server, which it accesses occasionally via the P-RAT link. Please refer to \cite{rajendran_journal2020} for more details.

\subsection{SCUBA UE Power Classes}\label{sec:powerclasses}
\begin{table}[t]
\centering
\caption{UE Specification for Different PCs}\label{table:power_classes}
\begin{tabular}{l|l|l|l}
\hline
Class   & Transmit Power      & \hfil Bandwidth         & \hfil Geo Regulation          \\ \hline\hline
PC1     & \hfil $14$ dBm            & $865-868$ MHz     & \hfil Europe                  \\ \hline
PC2     & \hfil$23$ dBm            & $902-928$ MHz      & \hfil North America           \\ \hline
\end{tabular}
\end{table}
Based on the regulations governing the use of different unlicensed frequency bands, we categorize SCUBA UEs into two power classes (PCs): PC1 and PC2. The physical layer specifications of the UEs in these two PCs are shown in Table~\ref{table:power_classes}, which are derived from the geo- and band-specific regulations~\cite{rajendran2020,rajendran_journal2020}.


\section{Synchronization for SCUBA}\label{sec:scubasync}
We begin by investigating if the existing synchronization strategies in the literature can be adopted to SCUBA. Wireless technologies that include a centralized BS, e.g., LTE, usually have the BS broadcasting a periodic synchronization signal (SS) that enables the UEs in the network to maintain sync by resynchronizing frequently. However, this is a power-expensive sync scheme to be directly employed in a distributed network like SCUBA. Technologies such as Bluetooth and Wi-Fi transmit a preamble~\cite{ieee802, bluetooth_spec} attached to every data packet which assists the receiver UE to synchronize to the sender and further decode the data. Alternatively, in-coverage LTE-D2D~\cite{lte_d2d_wp} utilizes the periodic synchronization provided by the LTE BS. However, the OOC and PC LTE-D2D UEs follow a synchronization approach similar to Bluetooth and Wi-Fi, where data is preceded by an SS. Since SCUBA is entirely designed by considering the underlying cellular P-RAT as the baseline protocol~\cite{rajendran_journal2020}, i.e., reusing radio and chipset of the underlying P-RAT, and since SCUBA network operates similar to the LTE-D2D network under different coverage scenarios, we consider LTE-D2D as a starting point for our SCUBA sync design. 

LTE-D2D uses a technique where a cell-edge or OOC SRC always transmits an SL synchronization signal (SLSS) before sending SL data~\cite{lte_d2d_wp}. The DST then synchronizes to the SRC using the SLSS before trying to decode the SL data. In LTE-D2D, the DST listens for SLSS within a time range, called SL synchronization window (SLSW)~\cite{ts_36331}, which can be configured to either $5$~ms or half of the normal cyclic prefix (CP). However, this method requires the UE to resynchronize frequently to maintain the synchronization error within the SLSW, resulting in a high power consumption. Furthermore, since the SLSW duration is fixed, it cannot be varied to either increase it to accommodate for larger sync errors or decrease to reduce power consumption for smaller sync errors. 

To quantify the extent of achieved synchronization, we define \textit{coarse} sync to be the condition when the time sync between the UEs, $\Delta_\sync$, follows
\begin{equation}\label{eq:coarse_sync_condition}
    t_{\text{CP}} \leq \Delta_\sync \leq t_{\text{SLSW}},
\end{equation}
where $t_{\text{CP}}$ is the cyclic prefix length and $t_{\text{SLSW}}$ is the length of the SLSW, both in time. Similarly, we define two UEs to be in \textit{fine} sync with each other when
\begin{equation}\label{eq:fine_sync_condition}
    \Delta_\sync <  t_{\text{CP}}.
\end{equation}
For successful SCUBA operation that ensures that an SRC can precisely compute the SL-PO of the DST for SL data transfer, it is essential to meet the condition~\eqref{eq:fine_sync_condition}.

With this backdrop, we analyze the potential adaptations of the LTE-D2D compatible synchronization methods and their applicability to SCUBA.
\subsection{Adaptations of Prior Art} \label{subsec:adaptation_priorart}
\subsubsection{Inter-cell synchronization}
UEs in COOS may be synchronized due to the inter-cell synchronization available between BSs of different cells~\cite{tr_38855}. However, this inter-cell synchronization is not always guaranteed~\cite{Tho2013RS}, and thus falls short of being a feasible solution for SCUBA synchronization.
\subsubsection{Sync range extension}
UEs that are OOC may often be able to synchronize to a nearby BS since they can decode SLSSs far beyond the boundary of the supported user-plane range~\cite{Tho2013RS}. However, this method requires significant time duration in the order of several seconds to synchronize given the low signal-to-noise ratio (SNR) conditions~\cite{Tho2013RS}. This in turn critically impacts the battery life of the UE.
\subsubsection{GNSS-sync}
Global navigation satellite system (GNSS) based synchronization is shown to achieve $ \Delta_\sync \gtrsim 40$~ns~\cite{gnss_gps}. While it meets the condition in~\eqref{eq:fine_sync_condition}, GNSS-based synchronization has several drawbacks. First, it is a power hungry scheme consuming several seconds for synchronization. Second, due to link budget constraints of commercial GNSS configurations, it can only work in outdoor environments. Furthermore, it also potentially requires additional hardware for GNSS signal processing. As a result, GNSS-based synchronization is unsuitable for low-cost and low-power cellular-IoT (C-IoT) applications. 
\subsubsection{TX-beacon method} 
LTE-D2D includes a provision to configure specific UEs, regardless of their cellular coverage, to periodically transmit SLSS, so as to extend synchronization to OOC UEs. These UEs act as SL transmit (TX) beacons which periodically transmit SLSS on pre-defined SFs followed by a broadcast message which includes the timing information such as system frame number (SFN)~\cite{lte_d2d_wp}. Such a method may be suitable for SCUBA. However, we show later in Section~\ref{sec:evaluation_results} that an adaptation of the beacon-based method, which uses a receiver (RX) beacon based strategy, is potentially more suitable for low-power C-IoT applications.  
\subsection{Proposed Solutions}
\subsubsection{Flexi-Sync Method}\label{sec:variable_sync_window}
We present our first solution, called flexi-sync method, for two types of pre-sync scenarios.
\begin{figure}[t]	
	\begin{center}
		\subfloat[]{\includegraphics[clip,width=0.40\columnwidth]{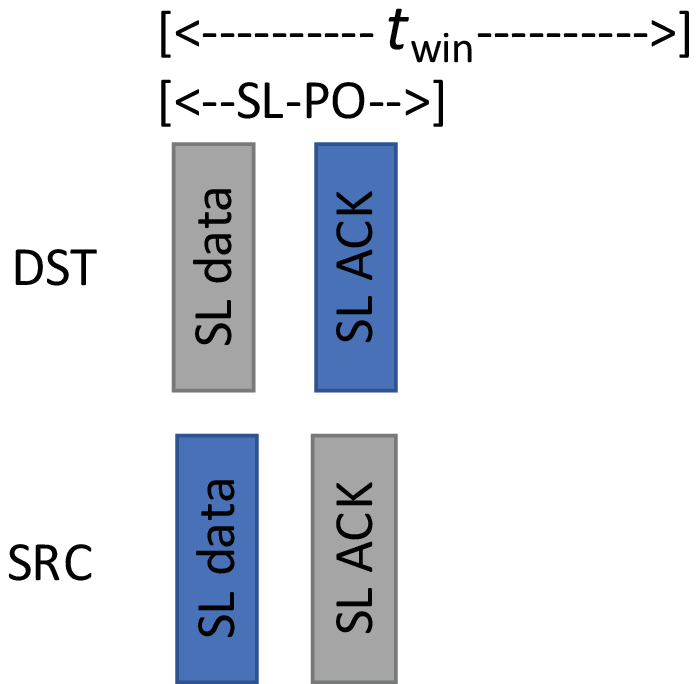}}
		\hspace{0.1\columnwidth}
		\subfloat[]{\includegraphics[clip,width=0.35\columnwidth]{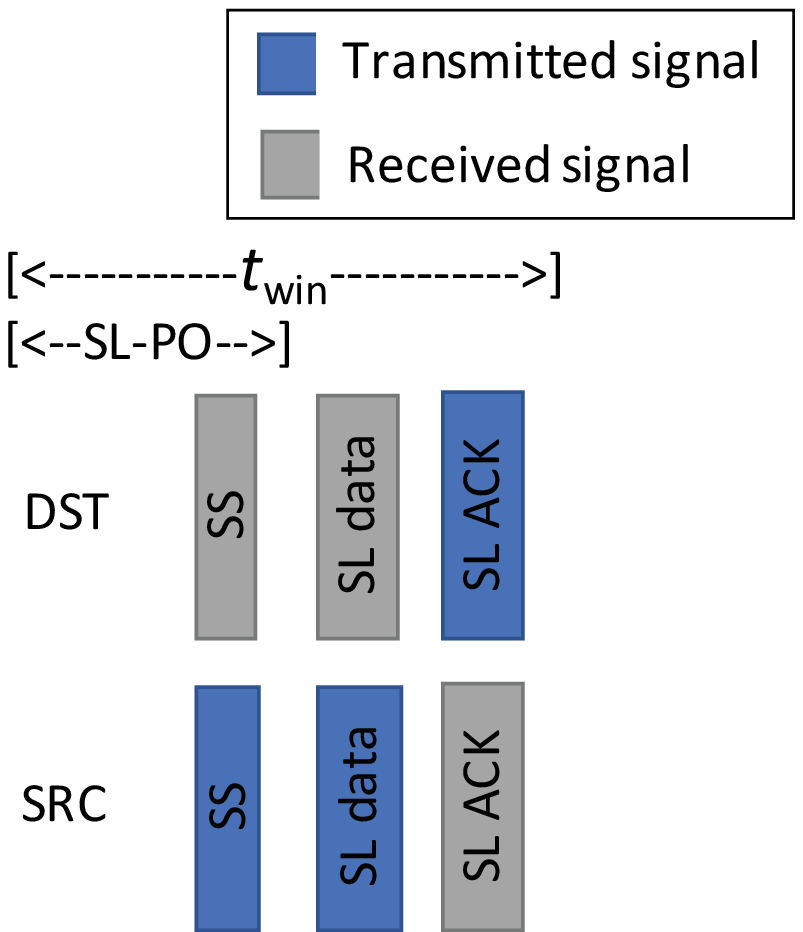}}
		\caption{SCUBA synchronization when (a) the SRC is in fine sync with the DST, and (b) when the SRC is not.}
		\label{fig:scuba_sync}
	\end{center}
	\vspace{-.2in}
\end{figure}

\textit{When UEs are coarsely synchronized}: 
Coarse synchronization can be achieved when the UEs are in the COOS and their respective BSs provide coarse synchronization between them. We borrow the idea of SLSW from LTE-D2D, and enforce SCUBA DST UEs to listen for a duration of $t_{\text{win}}$ around its SL-PO. However, unlike LTE-D2D, we let $t_{\text{win}}$ to be customizable by the UE application and the extent of coarse sync achievable. Accordingly, every SCUBA transmission is preceded by the SRC transmitting an SLSS on the SL-PO of the DST, which the SRC computes using its coarse timing. We replace the DST SL-PO with two listening occasions. The first is the conventional SL-PO where the DST listens for and decodes potential SL data from UEs that are already in fine sync with the DST. The second is a SCUBA sync window (SSW), which overlaps with the SL-PO, where the UE looks for and decodes an SLSS. We choose $t_{\text{win}}$ such that it accommodates the total of all types of timing synchronization errors, $\epsilon_\text{t}$, which is 
\begin{equation}\label{eq:total_sync_errors}
	\epsilon_\text{t} = \epsilon_\text{coarse} + \epsilon_\text{SRC} + \epsilon_\text{DST} + t_\text{d},
\end{equation}
where $\epsilon_\text{coarse}$ is the synchronization error resulting from coarse synchronization, $\epsilon_\text{SRC}$ and $\epsilon_\text{DST}$ are the SRC and DST crystal clock errors, respectively, and $t_\text{d}$ is the delay due to the time of flight between SRC and DST. By accommodating $t_\text{d}$ in~\eqref{eq:total_sync_errors}, our method ensures synchronization between UEs that are spaced arbitrarily far away from each other. The value of $\epsilon_\text{coarse}$ is dependent on the type of coarse synchronization achieved, as discussed in Section~\ref{subsec:adaptation_priorart}. The crystal clock errors, $\epsilon_\text{SRC}$ and $\epsilon_\text{SRC}$, are given by 
\begin{align}
	\epsilon_\text{SRC} &= x_\text{SRC}\cdot t_\text{coarse},\\
	\epsilon_\text{DST} &= x_\text{DST}\cdot t_\text{coarse},
\end{align}
respectively, where $x_\text{SRC}$ and $x_\text{DST}$ are the SRC and DST crystal clock inaccuracies per unit time respectively, and $t_\text{coarse}$ is the time elapsed since the latest coarse synchronization. The value of $t_\text{coarse}$ can either be set beforehand or varied dynamically. Therefore, the SRC needs to resynchronize close to when it transmits SL data to ensure that its error is within the defined maximum $\epsilon_\text{SRC}$. In Fig.~\ref{fig:scuba_sync}, we show different examples of SCUBA operation with our flexi-sync method. In Fig.~\ref{fig:scuba_sync}~(a), we illustrate the case where $\Delta_\sync < t_\text{CP}$. In such a case, a transmission of SLSS by the SRC is not required. Therefore, the SRC can directly transmit the SL data, which can be perfectly decoded at the DST. This option is suitable when the SRC is aware that it is in fine sync with the DST, either due to a previous successful SCUBA transmission in the near past or due to the prior knowledge that the SRC and DST are in HC. Thereby, we ensure power-optimized downward compatibility with HC SCUBA. The second case in Fig.~\ref{fig:scuba_sync}~(b) is the generic scenario where the SRC transmits an SLSS before its SL data transmission. The DST listens for SLSS during its SSW and upon reception of an SLSS, synchronizes itself to the SRC and then decodes the SL data that arrives in the following time slots.


\textit{When UEs are not coarsely synchronized}:
We propose an adaptation to the flexi-sync method that caters to the scenarios where UEs do not have coarse sync with each other. Such a situation may be encountered at cold-start, after losing coarse sync due to inactivity, or when COOS UEs do not have coarse synchronization. When the UEs are not even coarsely synchronized with each other, an SRC is not only unaware of the SL-PO of the DST but can also not reach its SSW. Therefore, we let an SRC UE transmit an SLSS followed by an SL timing request message (TimeREQ) until it receives an SL timing response (TimeRSP) from any UE in the listening neighborhood. 

\begin{figure}[t]	
	\begin{center}
		\subfloat[]{\includegraphics[clip,width=0.8\columnwidth]{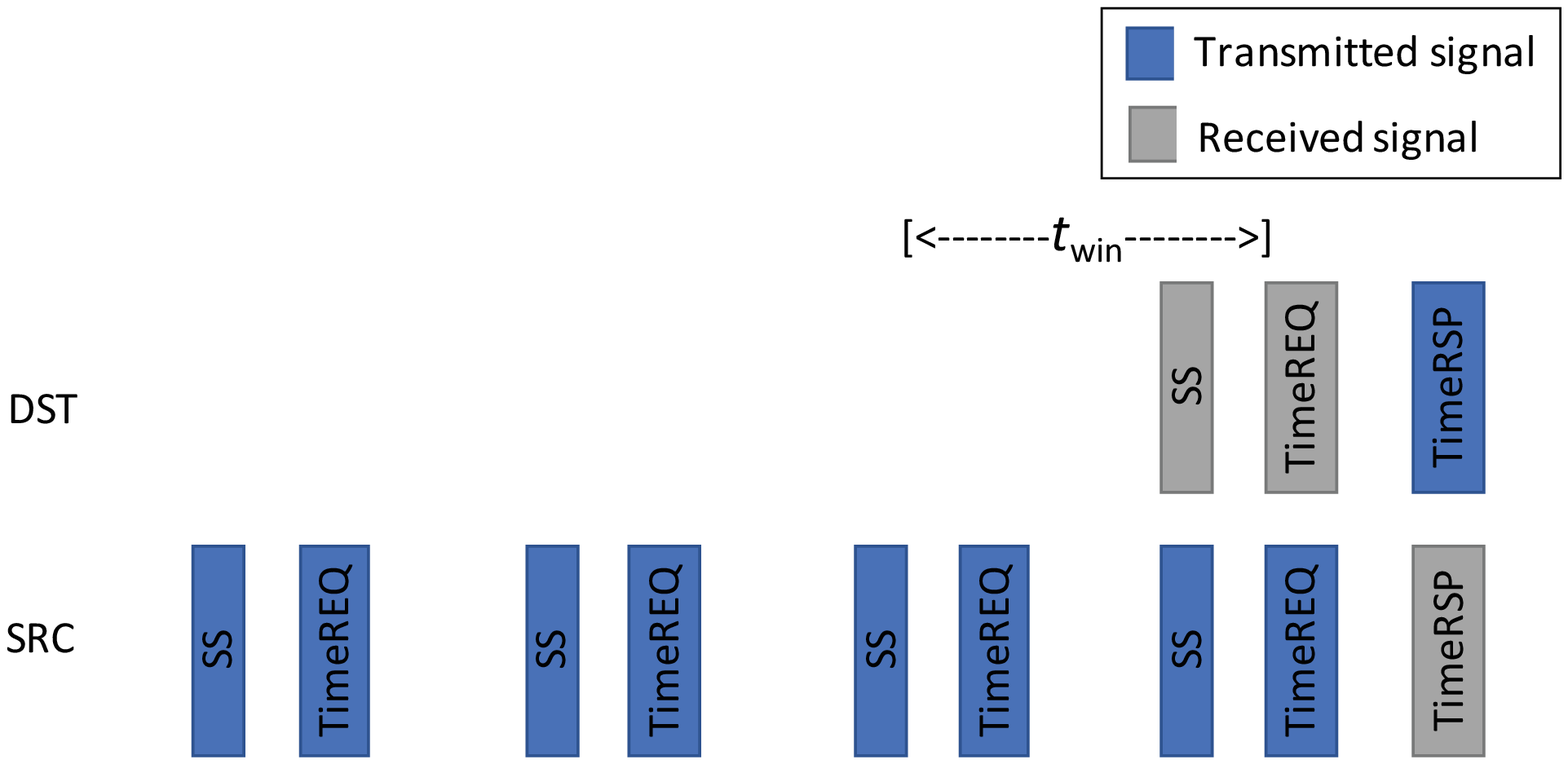}}
		\\
		\subfloat[]{\includegraphics[clip,width=0.8\columnwidth]{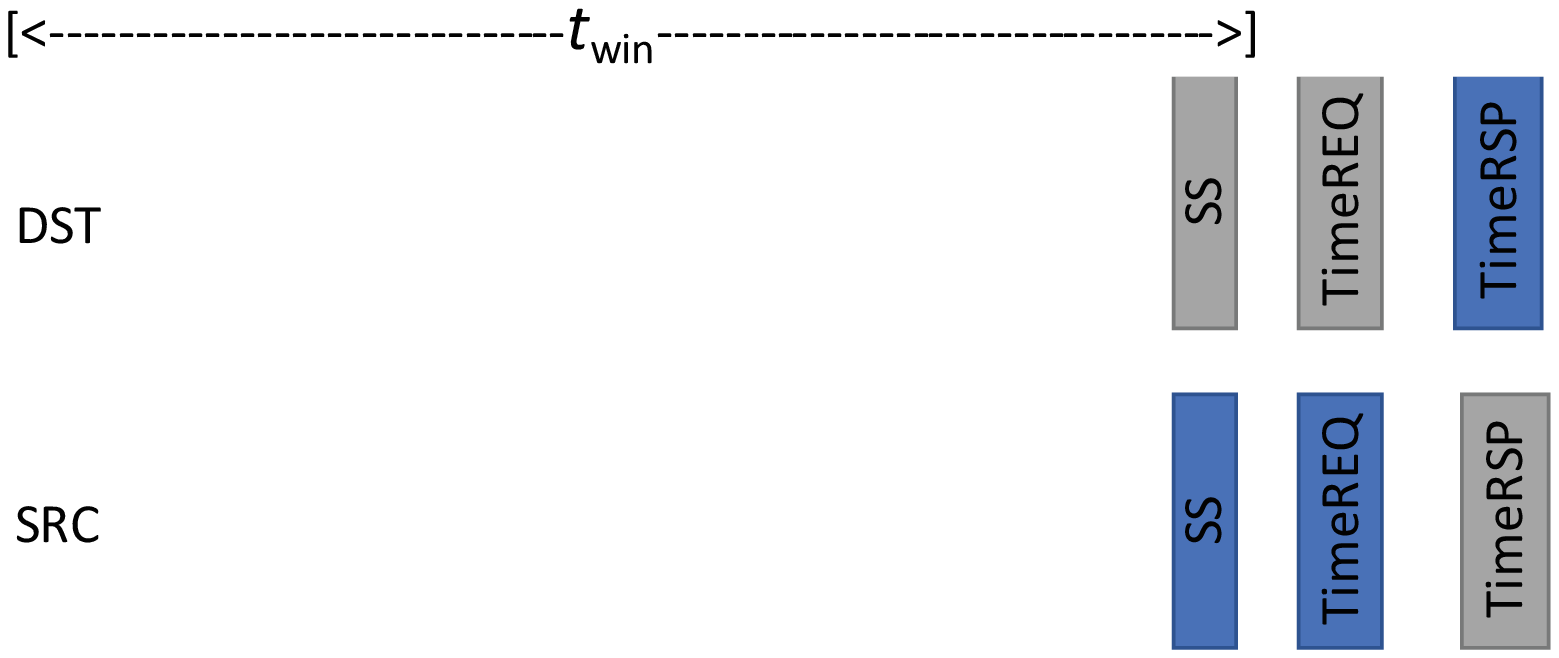}}
		\caption{Flexi-sync operation with TimeREQ and TimeRSP for (a) shorter $t_{\text{win}}$ with $N_\text{A} = 4$, and (b) extended $t_{\text{win}}$ with $N_\text{A}=1$.}
	    \label{fig:timereq}
	\end{center}
	\vspace{-0.2in}
\end{figure}

The number of attempts, $N_\text{A}$, required by the SRC to encounter an SSW of the DST is dependent on $t_{\text{win}}$. A longer $t_{\text{win}}$ requires a smaller $N_\text{A}$ and vice versa. The choice of $N_\text{A}$ and $t_{\text{win}}$ drives the power consumption at the SRC and DST, respectively. Fig.~\ref{fig:timereq}~(a) and Fig.~\ref{fig:timereq}~(b) show instances of varying $t_{\text{win}}$ to demonstrate the impact on $N_\text{A}$. Our method can also be modified to piggyback TimeREQ message with the SLSS to further reduce signaling by the SRC at the expense of increased decoding complexity at the DST. The trade-offs between the power consumption of an SRC and DST is typically chosen during system configuration based on the traffic type and battery life constraints at the UE.


\subsubsection{Sync Beacon Method}\label{sec:sync_beacon}
For our second proposed method, we borrow the beacon-based synchronization scheme from LTE-D2D. This technique is suitable for UEs which do not have a strict constraint on power consumption, e.g., UEs with large batteries or alternating current (AC) powered devices. Beacon based synchronization is also applicable for UEs with or without coarse synchronization.

We define two types of beacons, TX and RX beacons, which transmit and receive beacons, respectively. A TX beacon approach is borrowed directly from LTE-D2D standard~\cite{lte_d2d_wp}, where a SCUBA UE transmits beacons at regular intervals for synchronization by any other SCUBA UE. However, this method is inefficient for SCUBA UEs synchronizing to the beacons by listening for long SSW, especially for PC1 SCUBA UEs having comparable TX and RX powers. Instead, we propose an RX beacon method, where beacon UEs meant to provide synchronization to other SCUBA devices listen for a TimeREQ message that may be transmitted by a SCUBA UE intending to obtain timing information. This is similar to the TimeREQ method with the modification that the receiving UE is an RX beacon that is meant to serve the SCUBA network to provide timing information on demand. In Section~\ref{sec:evaluation_results}, we show that an RX beacon method provides superior energy efficiency when used with low transmit power devices, e.g., PC1 SCUBA UEs. It should be noted that the beacon devices are also SCUBA UEs which have their P-RATs operating in TDM manner with SCUBA, and hence beacon periodicity or continuity is not always guaranteed. In cases where RX beacons are incapable of listening for sync requests continuously or TX beacons are unable to send SLSS at frequent intervals, the synchronizing SCUBA UE may make multiple attempts to achieve a successful sync.

\subsection{Choosing Sync Methods}
Our proposed solutions are independent of each other and are capable of providing synchronization in all scenarios, including cold-start. While the beacon-based synchronization methods allow UEs to maintain sync by resynchronizing periodically with a beacon node that has no restrictions on power consumption, the flexi-sync method is a data-driven technique that requires UEs to sync only when exchanging SCUBA messages. However, the network may also choose to use a mixture of the two methods. For example, a network can use a beacon node with a large synchronization interval, with the option of also using flexi-sync between two UEs. This allows the UEs to use a smaller $t_\text{win}$ and still encounter low values of $N_\text{A}$ for beacon-based synchronization. This provides the network with greater flexibility in choosing latency and battery life trade-off.   

\subsection{SLSS Design}
The design of SLSS can be adopted directly from LTE-D2D as the signals largely serve the same purpose. However, we present two modifications in the following to adapt it for SCUBA applications.

\subsubsection{SCUBA Server Substitution}
The use of our proposed methods not only solves the issue of synchronization in UEs for OOC, PC, and COOS conditions, but also eliminates the need for a central SCUBA server. Legacy SCUBA relies on a central server to extract DST information such as DST UE ID and SL-DRX cycle values to compute SL-PO. On the other hand, using our synchronization methods, where an SRC-based synchronization is supported, eliminates the need for SRC to compute the SL-PO of the DST beforehand prior to initiating SCUBA transmission. Furthermore, the UE ID and the SL-PO of the DST is also embedded within the SLSS so that the SRC can use them for subsequent transmissions without contacting a central server. 

\subsubsection{Pseudo-unique SLSS}\label{subsubsec:unique_sync}
The SLSSs are meant to be received by a DST or an RX beacon in the SSW that is positioned around the SL-PO. SCUBA allocates SL-PO to be pseudo-unique by having them be dependent on the UE ID of the DST. However, while a $1$~ms long SL-PO, as defined in SCUBA, can be largely non-overlapping between DST UEs even in a crowded network, the SSW is considerably larger in time than the SL-PO. Therefore, the probability of inter-SSW overlap is higher, and so is the rate of SLSS collision. To counter this, we propose pseudo-uniqueness to be embedded within the SLSS using the DST UE ID, when known. This reduces the probability of false alarms of detecting an SLSS at the DST UE.


\section{Performance Analysis}\label{sec:analysis}
In this section, we analyze the power consumption for achieving synchronization using our proposed methods. 

\subsection{Flexi-Sync}
For the flexi-sync method, we consider the performance independently for SRC and DST UEs since they consume power asymmetrically. This allows us to prioritize the performance individually. For an SRC UE that requires $N_{\text{A}}$ attempts to achieve synchronization, the average power consumed is 
\begin{equation}\label{eq:power_ma_src}
	P_\text{SRC} = \frac{N_{\text{A}}}{T_\text{data}}\bigg(P_{\text{TX}}\big(t_{\text{SS}}+t_{\text{req}}\big)+P_{\text{RX}}t_{\text{rsp}}\bigg),
\end{equation}
where $P_{\text{TX}}$ is the SCUBA transmission power at the UE, $P_{\text{RX}}$ is the SCUBA reception power at the UE, $T_\text{data}$ is the mean inter-arrival time of SCUBA data, and $t_{\text{SS}}$, $t_{\text{req}}$, and $t_{\text{rsp}}$ are the time durations of SLSS, TimeREQ, and TimeRSP signals, respectively. 
The corresponding power consumption in the DST is given by
\begin{equation}
	P_\text{DST} = \frac{1}{T_\text{data}}\bigg(P_{\text{RX}}t_{\text{win, eff}}+P_{\text{TX}}t_{\text{rsp}}\bigg),\label{eq:power_ma_dst}
\end{equation}
where
\begin{equation}
	t_{\text{win, eff}}= \frac{t_{\text{win}}}{N_{\text{A}}} 
	\label{eq:power_wineff}
\end{equation}
is the effective reduced SSW when multiple attempts are performed.

\subsection{Resync Using TX and RX Beacons}
For the beacon method, we analyze the power consumption in a SCUBA UE synchronizing periodically to a TX or RX beacon. Since the beacons are generally AC powered devices which are not power-critical, we do not analyze the power consumption in them. The total power consumption in a SCUBA UE periodically synchronizing to an RX beacon is given by
\begin{equation}
	P_\text{sync, RXbeacon} = \frac{N_{\text{A}}}{T_\text{sync}}\bigg(P_{\text{TX}}\big(t_{\text{SS}}+t_{\text{req}}\big)+P_{\text{RX}}t_{\text{rsp}}\bigg),\label{eq:power_rxbeacon}
\end{equation}
where $T_\text{sync}$ is the sync interval of the UE, which corresponds to the maximum error allowed in the SCUBA system.
On the other hand, the power consumption in a SCUBA UE for synchronizing with a SCUBA TX beacon is given by 

\begin{equation}\label{eq:power_txbeacon}
	P_\text{sync, TXbeacon} = \frac{N_{\text{A}}}{T_\text{sync}}\big(P_{\text{RX}}t_{\text{win}}\big).
\end{equation}
\section{Numerical Results}\label{sec:evaluation_results}
Since C-IoT applications are not latency critical, and SCUBA is designed to function as an S-RAT only when UEs are free from P-RAT, we consider power consumption as the metric to evaluate the performance of our proposed methods. Unless otherwise specified in the following sections, we use the values shown in Table~\ref{table:evaluation_settings} for the evaluations, which are obtained from~\cite{lte_d2d_wp, ts_36101, ericsson_tdoc}. 

\begin{table}[]
\centering
\caption{Evaluation Settings}\label{table:evaluation_settings}
\begin{tabular}{c|c||c|c}
\hline
Parameter        & Value & Parameter        & Value         \\ \hline\hline
$P_{\text{TX}}$  & $100$~mW & $T_{\text{data}}$ & $2$~hours   \\ \hline
$P_{\text{RX}}$  & $80$~mW  & $t_{\text{win}}$  & $72$~ms   \\ \hline
$t_{\text{SS}}$  & $1$~ms   & $T_{\text{sync}}$ & $8.33$~min   \\ \hline
$t_{\text{req}}=t_{\text{rsp}}$ & $1$~ms   & $N_{\text{A}}$ & $1$   \\ \hline
\end{tabular}
\end{table}
\subsection{Flexi-Sync Method}
\begin{figure}[t]
	\centering
	{\includegraphics[width=0.5\columnwidth]{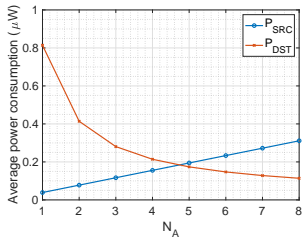}}
	\caption{Average power consumption in SRC and DST for SCUBA resync using multiple attempts.}
	\label{fig:results1} 
\end{figure}
We show the power consumption results of our flexi-sync solution in Fig.~\ref{fig:results1} for a range of values of $N_\text{A}$. We observe that while the power consumption in SRC increases linearly with increasing $N_\text{A}$, DST consumes lesser power with higher number of SRC TX attempts as it is required to listen for smaller durations of SSWs. Therefore, the results suggest that, if the SRCs in a SCUBA network are power-critical devices and the DSTs are not, for e.g., sensors reporting to a central controller node, lower number of sync attempts along with longer duration of SSW is preferable.

\subsection{Beacon-Based Method}
\begin{figure}[t]	
	\begin{center}
		\subfloat[]{\includegraphics[clip,width=0.49\columnwidth]{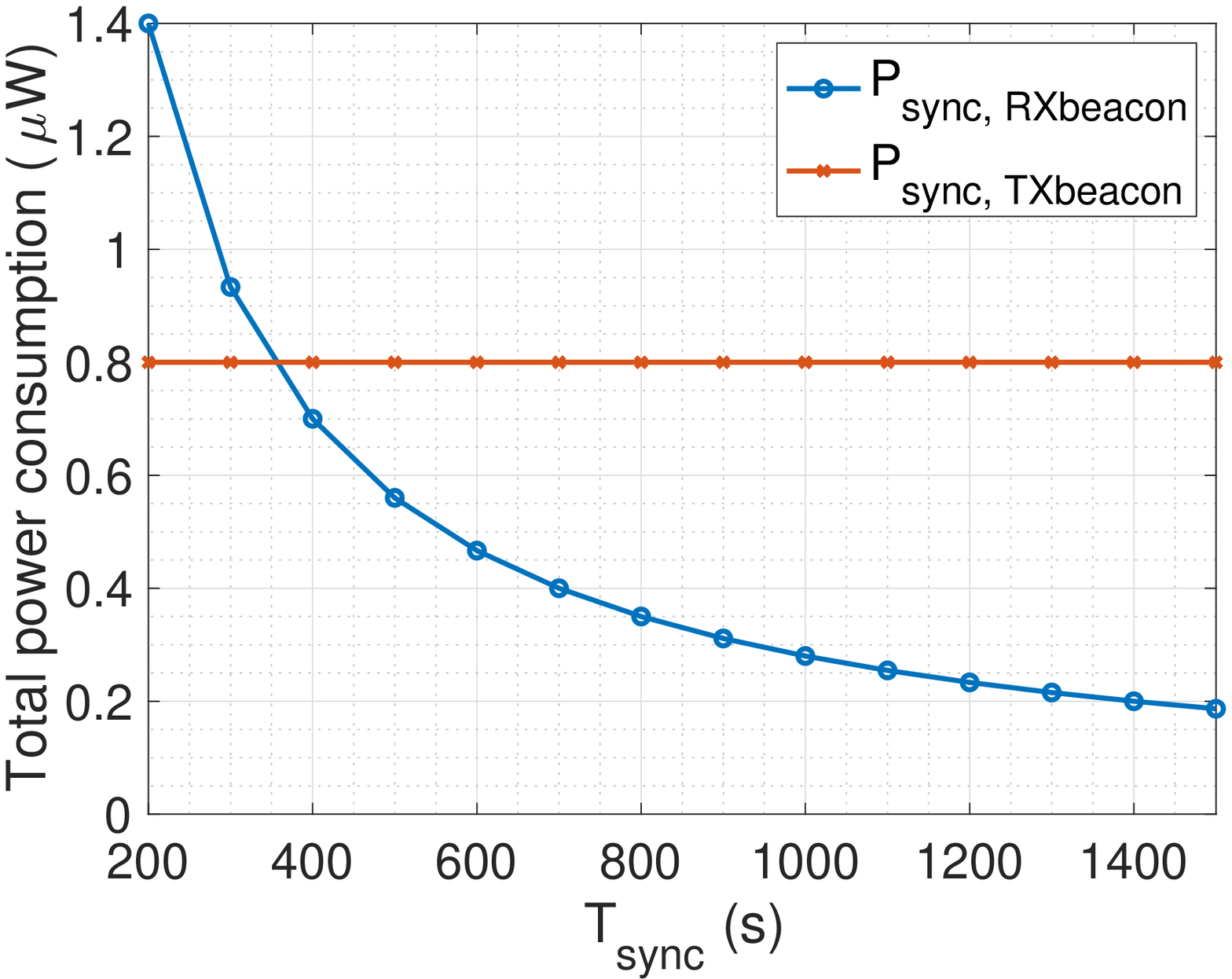}}
		\subfloat[]{\includegraphics[clip,width=0.51\columnwidth]{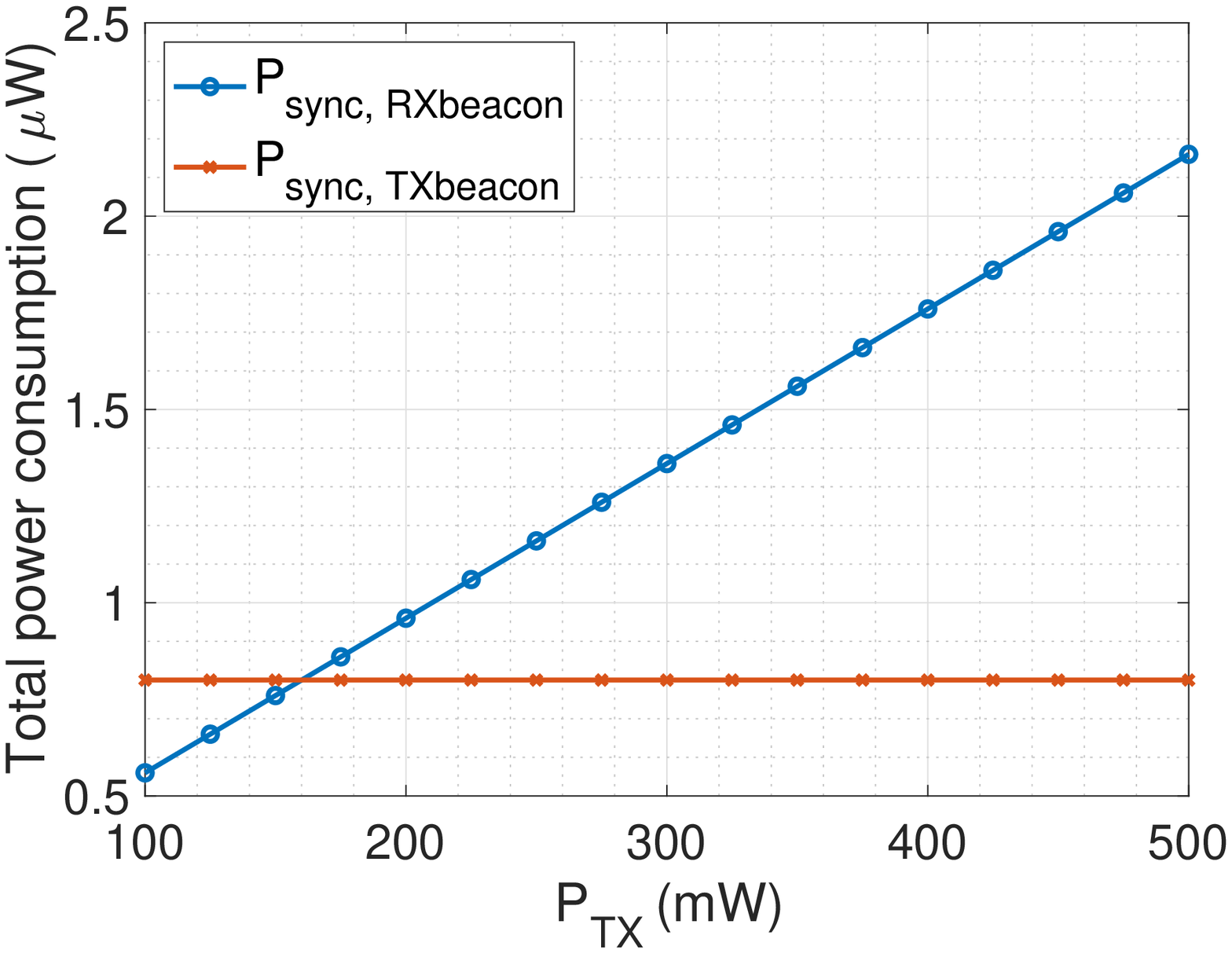}}
		\caption{Variation of total power consumption for the beacon-based synchronization method as a function of (a) sync interval and (b) transmit power.}
	    \label{fig:results2}
	\end{center}
	\vspace{-0.5cm}
\end{figure}

The variation of the total power consumption in a SCUBA UE for each of the beacon-based methods against a range of sync intervals is shown in Fig.~\ref{fig:results2}~(a). For the evaluation, we choose an optimal $t_{\text{win}}$ corresponding to each value of sync interval. For RX beacon based sync scheme, the power consumption in the SCUBA UE reduces with increasing values of sync interval. For the TX beacon based sync scheme, the power consumption in the UE remains constant regardless of the sync interval since $t_{\text{win}}$ is chosen optimally. For $T_{\text{sync}}>350~\text{s}$, SCUBA UE synchronizing to an RX beacon has lower power consumption compared to that of TX beacon. This suggests that RX beacon based synchronization is preferable when the sync interval in the SCUBA network is large.

Next, we present the evaluation results to show the impact of transmit power on the total power consumption in Fig.~\ref{fig:results2}~(b). For this evaluation, we choose $T_{\text{sync}}=8.33~\text{min}$ which corresponds to the time during which the accumulated sync error reaches $5$~ms, equivalent to the allowed SLSW in LTE-D2D. The power consumption in SCUBA UE for RX beacon based method increases with increasing values of transmission power. However, since there are no transmissions involved in the SCUBA UE when synchronizing to a TX beacon, the power consumption remains constant for the TX beacon based sync scheme. The power consumption traces intersect at $P_\text{TX}=160$~mW, which corresponds to SCUBA effective radiated power (ERP) of $45$~mW ($16.53$~dBm) for $45\%$ power amplifier efficiency and $60$~mW power consumption in the support circuitry~\cite{ericsson_tdoc}. We call the intersection point of power consumption traces of RX and TX beacon based methods as beacon power threshold ($P_\text{BTh}$), which plays a crucial role in choosing the appropriate sync scheme for a SCUBA network. The results suggest that RX beacon based synchronization scheme is preferable for SCUBA UEs belonging to those power classes transmitting at power lower than $P_\text{BTh}$. It should also be noted that the value of $P_\text{BTh}$ will be higher for higher values of $T_{\text{sync}}$. 

\subsection{Battery Life}
Finally, we present numerical values of the impact of our proposed solutions on the bottom-line metric for C-IoT device performance of battery life. To this end, we integrate our power consumption numbers with the operating power of native SCUBA protocol from~\cite{rajendran_journal2020}. Since availability of non power-critical beacon devices is not always guaranteed in a SCUBA network, we use flexi-sync method for the battery life analysis. For an MTC traffic model~\cite{tr_36888}, the battery life of a legacy SCUBA UE that coexists with LTE-M as P-RAT and utilizes the LTE-M network sync for SCUBA, is $328.3$~days~\cite{rajendran_journal2020}.
For a SCUBA device that only uses the flexi-sync method for synchronization, the battery life is $328.1$~days with $P_\text{TX}=100$~mW and $N_\text{A}=4$. The flexi-sync based SCUBA sync, which enables UE synchronization in all types of cellular coverage scenarios, thus results in less than $0.1\%$ reduction in battery life compared to legacy SCUBA condition that uses the LTE-M network sync. Thus, our method extends the operation range and provides seamless operation in all coverage scenarios for mobile C-IoT devices only at a cost of negligible reduction in battery life.    

\section{Conclusion}\label{sec:conclusion}
We have presented synchronization schemes for SCUBA devices to enable seamless D2D communication in mobile C-IoT UEs across all types of cellular coverage scenarios. Our solutions also ensure that time-of-flight does not limit the communication range of SCUBA devices. We provide flexible solutions that are adaptable based on UE hardware limitations and unlicensed band usage regulations. Numerical results showed that our proposed low-power solutions can achieve synchronization with a negligible impact on UE battery life. Comparison of other application-specific performance indicators for tailored network topologies is a straightforward extension of our work. Although our proposed methods are intended for SCUBA, they are adaptable to other types of D2D communication technologies, such as new radio (NR) sidelink. Synchronized SCUBA also strengthens the potential of being integrated into the MulteFire standard, which provides solutions for operating cellular communications on unlicensed bands.

\bibliographystyle{IEEEtran}
\bibliography{References}
\end{document}